\documentclass[aps,preprint]{revtex4}%
\usepackage{amsfonts}
\usepackage{amsmath}
\usepackage{amssymb}
\usepackage{graphicx}%
\begin{document}
\preprint{ }
\title{Inequalities for Light Nuclei in the Wigner Symmetry Limit}
\author{Jiunn-Wei Chen}
\affiliation{Department of Physics and National Center for Theoretical Sciences at Taipei,
National Taiwan University, Taipei 10617, Taiwan}
\author{Dean Lee}
\affiliation{Department of Physics, North Carolina State University, Raleigh, NC 27695, USA}
\author{Thomas Schaefer}
\affiliation{Department of Physics, North Carolina State University, Raleigh, NC 27695,
USA\linebreak RIKEN-BNL Research Center, Brookhaven National Laboratory,
Upton, NY 11973, USA}

\begin{abstract}
Using effective field theory we derive inequalities for light nuclei in the
Wigner symmetry limit. \ This is the limit where isospin and spin degrees of
freedom can be interchanged. \ We prove that the energy of any three-nucleon
state is bounded below by the average energy of the lowest two-nucleon and
four-nucleon states. \ We show how this is modified by lowest-order terms
breaking Wigner symmetry and prove general energy convexity results for
$SU(N)$. \ We also discuss the inclusion of Wigner-symmetric three and
four-nucleon force terms.

\end{abstract}
\maketitle
\preprint{ }

Weinberg was the first to apply effective field theory to the nucleon-nucleon
interaction \cite{Weinberg:1990rz}. \ Since then effective field theory
methods for nuclear physics have been further developed and applied
successfully to the two and three-nucleon system \cite{Ordonez:1996rz}%
\cite{Kaplan:1996xu}\cite{Kaplan:1998we}\cite{Beane:2000fx}%
\cite{Bedaque:2002mn}. \ Recently though there has also been progress in
applying effective field theory to many-body nuclear physics
\cite{Muller:1999cp}\cite{Chen:2003vy}\cite{Lee:2004si}\cite{Abe:2003fz}.

The traditional approach to many-body nuclear physics is based on two and
three-body potential models. \ Although this approach has been very successful
in describing the ground-state properties of light nuclei and neutron drops,
the effective field theory approach offers several important advantages.
\ First of all effective field theory provides a direct connection with
quantum chromodynamics (QCD). \ It explains the form of the interactions and
how the strength of the interaction changes with the cutoff scale. \ Also the
calculations can be systematically improved and one can estimate the errors
due to contributions that have been neglected. \ Furthermore it has been
discovered that the effective Lagrangian is often amenable to analysis so that
non-perturbative results can be proven without even any numerical simulations.
\ Effective field theory was used in \cite{Lee:2004ze} to prove inequalities
for the correlations of two-nucleon operators in symmetric nuclear matter.
\ In \cite{Lee:2004hc} upper bounds were proven for pressure in
isospin-asymmetric nuclear matter and neutron matter in a magnetic field.

The idea of proving rigorous inequalities from Euclidean functional integrals
is not new. In fact the connection between effective field theory inequalities
and nuclear lattice simulations is much the same as the connection between QCD
inequalities \cite{Weingarten:1983uj}\cite{Vafa:1983tf}\cite{Vafa:1984xg}%
\cite{Vafa:1984xh}\cite{Witten:1983ut}\cite{Nussinov:1983vh}%
\cite{Nussinov:1999sx}\cite{Nussinov:2003uj}\cite{Cohen:2003ut} and lattice
QCD. \ In this letter we first confirm the work of \cite{Mehen:1999qs} by
showing that the physics of low-energy symmetric nuclear matter is close to
the Wigner limit, where the isospin and spin degrees of freedom can be
interchanged. \ In this limit $SU(2)\times SU(2)$ spin-isospin symmetry is
elevated to an $SU(4)$ symmetry. \ We prove that the energy of any
three-nucleon state is bounded below by the average energy of the lowest
two-nucleon and lowest four-nucleon states. \ We show how this is modified by
lowest-order terms breaking Wigner symmetry and prove general energy convexity
results for $SU(N)$. \ We also discuss the inclusion of Wigner-symmetric three
and four-nucleon force terms. \ Much of the physics we describe is universal
and appears in other systems such as trapped Fermi gases near a Feshbach
resonance \cite{Kinast:2004}\cite{Regal:2004}. \ Our general result for
$SU(N)$ will be interesting if and when one is able to trap four or more
degenerate fermionic states.

Let $N$ represent the nucleon fields. \ We use $\vec{\tau}$ to represent Pauli
matrices acting in isospin space and $\vec{\sigma}$ to represent Pauli
matrices acting in spin space. \ We assume exact isospin symmetry.\ \ In the
non-relativistic limit and below the threshold for pion production, we can
write the lowest order terms in the effective Lagrangian as%
\begin{align}
\mathcal{L}  &  =\bar{N}[i\partial_{0}+\tfrac{\vec{\nabla}^{2}}{2m_{N}}%
-(m_{N}^{0}-\mu)]N-\tfrac{1}{2}C_{S}\bar{N}N\bar{N}N\nonumber\label{U}\\
&  -\tfrac{1}{2}C_{odd}\left[  \bar{N}\vec{\sigma}N\cdot\bar{N}\vec{\sigma
}N-\bar{N}\vec{\tau}N\cdot\bar{N}\vec{\tau}N\right]  .
\end{align}
We have neglected three-nucleon terms for now but will consider them later.
\ We have written the Lagrangian so that the operator multiplying $C_{odd}$
flips sign under the exchange of isospin and spin degrees of freedom.

We now calculate $C_{S}$ and $C_{odd}$ on a spatial lattice for various
lattice spacings $a_{lattice}$. \ We sum all nucleon-nucleon scattering bubble
diagrams on the lattice, locate the pole in the scattering amplitude, and
compare with L\"{u}scher's formula for energy levels in a finite periodic box
\cite{Luscher:1986pf}\cite{Beane:2003da}. \ In Table 1 we have computed these
coefficients.%
\[%
\genfrac{}{}{0pt}{0}{\text{Table 1: Contact potential coefficients }}{%
\begin{tabular}
[c]{|l|l|l|}\hline
$a_{lattice}^{-1}$(MeV) & $C_{S}\text{ (MeV}^{-2}\text{)}$ & $C_{odd}\text{
(MeV}^{-2}\text{)}$\\\hline
$20$ & $-3.40\times10^{-4}$ & $-3.8\times10^{-5}$\\\hline
$40$ & $-1.20\times10^{-4}$ & $-6.\times10^{-6}$\\\hline
$60$ & $-7.70\times10^{-5}$ & $-2.4\times10^{-6}$\\\hline
$80$ & $-5.60\times10^{-5}$ & $-1.3\times10^{-6}$\\\hline
$100$ & $-4.40\times10^{-5}$ & $-8.\times10^{-6}$\\\hline
\end{tabular}
\ \ }%
\]
\ As noted in \cite{Bedaque:1998kg}\cite{Bedaque:1999ve}, deeply-bound Efimov
states will begin to appear if we make $a_{lattice}^{-1}$ too large. \ In
reality this is not much of a restriction since we should not let
$a_{lattice}^{-1}$ exceed the chiral symmetry breaking scale (or even $m_{\pi
}$ for the pionless theory).

We see that $C_{odd}$ is much smaller in magnitude than $C_{S}$. \ In the
limit $C_{odd}\rightarrow0$ the $SU(2)\times SU(2)$ spin-isospin symmetry is
elevated to an $SU(4)$ symmetry. \ This symmetry was first studied by Wigner
\cite{Wigner:1939a}\cite{Wigner:1939b}\cite{Wigner:1939c}, and arises
naturally in the limit of large number of colors, $N_{c}$ \cite{Kaplan:1995yg}%
\cite{Kaplan:1996rk}. \ Although the $^{1}S_{0}$ and $^{3}S_{1}$ scattering
lengths are quite different, the fact that both scattering lengths are large
suggests we are close to the Wigner limit \cite{Mehen:1999qs}.

When $C_{odd}=0$ the grand canonical partition function is given by%
\begin{equation}
Z_{G}=\int DND\bar{N}\exp\left(  -S_{E}\right)  =\int DND\bar{N}\exp\left(
\int d^{4}x\,\mathcal{L}_{E}\right)  ,
\end{equation}
where
\begin{equation}
\mathcal{L}_{E}=-\bar{N}[\partial_{4}-\tfrac{\vec{\nabla}^{2}}{2m_{N}}%
+(m_{N}^{0}-\mu)]N-\tfrac{1}{2}C_{S}\bar{N}N\bar{N}N.
\end{equation}
Using Hubbard-Stratonovich transformations,\ we can rewrite $Z_{G}$ as%
\begin{equation}
Z_{G}\propto\int DND\bar{N}Df\exp\left(  \int d^{4}x\,\mathcal{L}_{E}%
^{f}\right)  ,
\end{equation}
where%
\begin{equation}
\mathcal{L}_{E}^{f}=-\bar{N}[\partial_{4}-\tfrac{\vec{\nabla}^{2}}{2m_{N}%
}+(m_{N}^{0}-\mu)]N+C_{S}f\bar{N}N+\tfrac{1}{2}C_{S}f^{2}.
\end{equation}
Since $C_{S}<0$, the $f$ integration is convergent.

Let $M$ be the nucleon matrix. \ $M$ has the block diagonal form,%
\begin{equation}
M=M_{\text{block}}\oplus M_{\text{block}}\oplus M_{\text{block}}\oplus
M_{\text{block}},
\end{equation}
where we have one block for each of the four nucleon states and%
\begin{equation}
M_{\text{block}}=-\left[  \partial_{4}-\tfrac{\vec{\nabla}^{2}}{2m_{N}}%
+(m_{N}^{0}-\mu)\right]  +C_{S}f.
\end{equation}
We note that $M$ is real valued and therefore $\det M\geq0$.

Consider the two-nucleon operator $A_{2}(x)=[N]_{i}[N]_{j}(x)$, where $i\neq
j$ are indices for two different nucleon states. \ The two-point correlation
function for $A_{2}$ is%
\begin{equation}
\ \left\langle A_{2}(x)A_{2}^{\dagger}(0)\right\rangle _{\mu,T}=\left\langle
[N]_{i}[N]_{j}(x)\;[N^{\ast}]_{j}[N^{\ast}]_{i}(0)\right\rangle _{\mu,T}.
\end{equation}
Using our Euclidean functional integral representation, we have%
\begin{equation}
\left\langle A_{2}(x)A_{2}^{\dagger}(0)\right\rangle _{\mu,T}=\int
D\Theta\;\left[  M_{\text{block}}^{-1}(x,0)\right]  ^{2}%
\end{equation}
where $D\Theta$ is the positive normalized measure defined by%
\begin{equation}
D\Theta=\dfrac{Df\;\det M\exp\left(  \tfrac{1}{2}C_{S}\int d^{4}%
x\,f^{2}\right)  }{\int Df\;\det M\exp\left(  \tfrac{1}{2}C_{S}\int
d^{4}x\,f^{2}\right)  }.
\end{equation}
We note that since $M_{\text{block}}$ is real valued, $M_{\text{block}}^{-1}$
is also real valued.

Next we consider the three-nucleon and four-nucleon operators $A_{3}%
(x)=[N]_{i}[N]_{j}[N]_{k}(x)$ and $A_{4}(x)=[N]_{i}[N]_{j}[N]_{k}[N]_{l}(x)$,
where $i,j,k,l$ are all distinct. \ We have%
\begin{align}
\left\langle A_{3}(x)A_{3}^{\dagger}(0)\right\rangle _{\mu,T}  &  =\int
D\Theta\;\left[  M_{\text{block}}^{-1}(x,0)\right]  ^{3},\\
\left\langle A_{4}(x)A_{4}^{\dagger}(0)\right\rangle _{\mu,T}  &  =\int
D\Theta\;\left[  M_{\text{block}}^{-1}(x,0)\right]  ^{4}.
\end{align}
We note that
\begin{align}
\int D\Theta\;\left\vert M_{\text{block}}^{-1}(x,0)\right\vert ^{3}  &  =\int
D\Theta\;\left\vert M_{\text{block}}^{-1}(x,0)\right\vert \left[
M_{\text{block}}^{-1}(x,0)\right]  ^{2}\nonumber\\
&  \leq\sqrt{\int D\Theta\;\left[  M_{\text{block}}^{-1}(x,0)\right]  ^{2}%
}\sqrt{\int D\Theta\;\left[  M_{\text{block}}^{-1}(x,0)\right]  ^{4}},
\end{align}
where the second line is from the Cauchy-Schwarz inequality. \ Therefore%
\begin{equation}
\left\vert \left\langle A_{3}(x)A_{3}^{\dagger}(0)\right\rangle _{\mu
,T}\right\vert \leq\sqrt{\left\langle A_{2}(x)A_{2}^{\dagger}(0)\right\rangle
_{\mu,T}\left\langle A_{4}(x)A_{4}^{\dagger}(0)\right\rangle _{\mu,T}}.
\end{equation}
\qquad

Let $E_{A_{2}}$ be the energy of the lowest state that couples to $A_{2}$, and
$E_{A_{4}}$ be the energy of the lowest state that couples to $A_{4}$.
\ Taking the limit $x\rightarrow\infty$ in the temporal direction we conclude
that any state with the quantum numbers of $A_{3}$ must have energy less than
the average of $E_{A_{2}}$ and $E_{A_{4}},$%
\begin{equation}
E_{A_{3}}\geq\tfrac{1}{2}\left[  E_{A_{2}}+E_{A_{4}}\right]  . \label{energy}%
\end{equation}
Taking the limit $x\rightarrow\infty$ in any spatial direction we have that
the inverse correlation length for $A_{3}$ must be greater than the average of
the inverse correlation lengths for $A_{2}$ and $A_{4}$,%
\begin{equation}
\xi_{A_{3}}^{-1}\geq\tfrac{1}{2}\left[  \xi_{A_{2}}^{-1}+\xi_{A_{4}}%
^{-1}\right]  . \label{corrlength}%
\end{equation}
Using arguments similar to those in \cite{Lee:2004ze},\ we can show that the
inequalities (\ref{energy}) and (\ref{corrlength}) hold for a general three
nucleon operator%
\begin{equation}
A_{3}(x)=\int_{\Omega}d^{4}x_{1}d^{4}x_{2}a_{ijk}(x_{1},x_{2})[N]_{i}%
(x+x_{1})[N]_{j}(x+x_{2})[N]_{k}(x+x_{3}),
\end{equation}
so long as $\Omega$ is bounded.

In the real world $C_{odd}$ is small but nonzero. \ We can measure the shift
in the energy of a given state $\left\vert A\right\rangle $ using first-order
perturbation theory, $\Delta E_{A}=\left\langle A\right\vert H^{\prime
}\left\vert A\right\rangle $, where
\begin{equation}
H^{\prime}=\tfrac{1}{2}C_{odd}\int d^{3}\vec{x}\;\left[  \bar{N}\vec{\sigma
}N\cdot\bar{N}\vec{\sigma}N-\bar{N}\vec{\tau}N\cdot\bar{N}\vec{\tau}N\right]
.
\end{equation}

Let us first consider two-nucleon states in an $S$-wave. \ In terms of $SU(4)$
representations, we decompose the tensor product of two fundamental $4$
dimensional representations, $4\otimes4=10\oplus6$. The $6$ dimensional
representation is antisymmetric, and the spin and isospin representations must
be $6=(1,0)\oplus(0,1)$. \ The spin triplet with isospin singlet corresponds
with the deuteron, $D$, while the spin singlet with isospin triplet
corresponds with the nearly bound $^{1}S_{0}$ states. \ A similar analysis for
$S$-wave three-nucleon states gives us one antisymmetric $\bar{4}$
representation with spin-isospin content $\bar{4}=(\tfrac{1}{2},\tfrac{1}{2}%
)$. \ This multiplet corresponds with the triton$,T$, and Helium-3. \ There is
only one antisymmetric $S$-wave four-nucleon state. \ It is a therefore spin
singlet and isospin singlet and corresponds with Helium-4$.$

Under a transformation that interchanges spin and isospin degrees of freedom,
$\left\vert ^{4}\text{He}\right\rangle $ is mapped into itself, possibly with
a minus sign. \ However $H^{\prime}$ is odd under this transformation and
therefore, $\left\langle ^{4}\text{He}\right\vert H^{\prime}\left\vert
^{4}\text{He}\right\rangle =0.$ \ Under the interchange of spin and isospin,
the spin-up $^{3}$He state and spin-down $T$ state are also mapped into
themselves, possibly with minus signs. \ We conclude that $\left\langle
^{3}\text{He}\right\vert H^{\prime}\left\vert ^{3}\text{He}\right\rangle
=\left\langle T\right\vert H^{\prime}\left\vert T\right\rangle =0$. \ Under
the interchange of spin and isospin, the $D$ states and $^{1}S_{0}$ states
interchange with each other, again possibly with minus signs. \ Therefore
$\left\langle D\right\vert H^{\prime}\left\vert D\right\rangle
=-\,\left\langle ^{1}S_{0}\right\vert H^{\prime}\left\vert ^{1}S_{0}%
\right\rangle $. \ We can now adjust for the first-order energy corrections
due to $H^{\prime}$,%
\begin{equation}
E_{^{3}\text{He}},E_{T}\geq\tfrac{1}{2}\left[  \tfrac{1}{2}\left(
E_{D}+E_{^{1}S_{0}}\right)  +E_{^{4}\text{He}}\right]  .
\label{physicalresult}%
\end{equation}
The physical binding energies are shown in Table 2 \cite{Audi:1995dz}.%
\[%
\genfrac{}{}{0pt}{0}{\text{Table 2: \ Binding energies for light nuclides}}{%
\begin{tabular}
[c]{|l|l|}\hline
$^{1}S_{0}$ & $\sim0\text{ MeV (nearly bound)}$\\\hline
$D$ & $-2.224\text{ MeV}$\\\hline
$^{3}\text{He}$ & $-7.718\text{ MeV}$\\\hline
$T$ & $-8.481\text{ MeV}$\\\hline
$^{4}\text{He}$ & $-28.296\text{ MeV}$\\\hline
\end{tabular}
\ }%
\]
Plugging these values into (\ref{physicalresult}), we find that the inequality
is satisfied, $-7.7$ MeV$,-8.5$ MeV$\geq-14.7$ MeV. \ An analogous relation
can be derived for the inverse correlation lengths,%

\begin{equation}
\xi_{^{3}\text{He}}^{-1},\xi_{T}^{-1}\geq\tfrac{1}{2}\left[  \tfrac{1}%
{2}\left(  \xi_{D}^{-1}+\xi_{^{1}S_{0}}^{-1}\right)  +\xi_{^{4}\text{He}}%
^{-1}\right]  .
\end{equation}
We stress that all of these inequalities hold in symmetric nuclear matter at
any density or temperature where the effective theory description is valid.
\ From here on we will consider only energy inequalities since the inverse
correlation length inequalities are completely analogous.

We now generalize the results (\ref{energy}) and (\ref{corrlength}) to the
case where the Wigner symmetry is an $SU(N)$ symmetry, for arbitrary $N$.
\ Let $n_{small},n_{big},$ and $n$ be any integers such that $0\leq
2n_{small}<n<2n_{big}\leq N$. \ Then
\begin{equation}
E_{n}\geq\frac{n-2n_{small}}{2n_{big}-2n_{small}}E_{2n_{big}}+\frac
{2n_{big}-n}{2n_{big}-2n_{small}}E_{2n_{small}}. \label{generalenergy}%
\end{equation}
This inequality is a statement of convexity of the energy as a function of
nucleon number, with the additional requirements that the number of nucleons
is less than or equal to $N$ and the two endpoints have an even number of
nucleons. \ The proof of the inequality is straightforward. \ We can write%
\begin{equation}
\int D\Theta\;\left\vert M_{\text{block}}^{-1}(x,0)\right\vert ^{n}=\int
D\Theta\;\left\vert M_{\text{block}}^{-1}(x,0)\right\vert ^{z}\left\vert
M_{\text{block}}^{-1}(x,0)\right\vert ^{n-z}, \label{holder}%
\end{equation}
Applying the H\"{o}lder inequality to the right-hand side, one finds the upper
bound (\ref{generalenergy}).

Taking $N=4$ and setting $2n_{small}=2,$ $n=3$, $2n_{big}=4,$ we recover
(\ref{energy})$.$ \ If however we let $2n_{small}=0,$ $n=3$, $2n_{big}=4$, we
get $E_{3}\geq\tfrac{3}{4}E_{4}$. \ There are no first order $H^{\prime}$
corrections in this case, and we see that in the real world this inequality is
satisfied, $-7.7$ MeV$,-8.5$ MeV $\geq-21.2$ MeV. \ Setting $2n_{small}=0,$
$n=2$, $2n_{big}=4,$ we get $E_{2}\geq\tfrac{1}{2}E_{4}$. \ With first order
$H^{\prime}$ corrections we have%
\begin{equation}
\tfrac{1}{2}\left(  E_{D}+E_{^{1}S_{0}}\right)  \geq\tfrac{1}{2}%
E_{^{4}\text{He}},
\end{equation}
and this is also satisfied, $-1.1$ MeV $\geq-14.1$ MeV.

Up to this point we have ignored three and four-nucleon forces$.$ \ It has
been shown that the dominant three-nucleon force is Wigner-symmetric
\cite{Bedaque:1998kg}\cite{Bedaque:1999ve}. \ We now show that introducing
Wigner-symmetric three and four-nucleon forces do not spoil positivity of the
Euclidean functional integral so long as the three-nucleon force is not too
strong and the four-nucleon force is not too repulsive. \ We want to find a
Hubbard-Stratonovich transformation that reproduces a contribution to the
action of the form,%
\begin{equation}
\prod_{x}\exp\left[  c_{2}[\bar{N}N(x)]^{2}+c_{3}[\bar{N}N(x)]^{3}+c_{4}%
[\bar{N}N(x)]^{4}\right]  .
\end{equation}
Let us just concentrate on what happens at a single point $x,$ and in our
notation we suppress writing the $x$ explicitly. \ We note that $\bar{N}N$
raised to any power greater than $4$ mush vanish. \ So we have
\begin{align}
&  \exp\left[  c_{2}\left(  \bar{N}N\right)  ^{2}+c_{3}\left(  \bar
{N}N\right)  ^{3}+c_{4}\left(  \bar{N}N\right)  ^{4}\right]  \nonumber\\
&  =a_{0}+a_{1}\bar{N}N+\tfrac{a_{2}}{2!}\left(  \bar{N}N\right)  ^{2}%
+\tfrac{a_{3}}{3!}\left(  \bar{N}N\right)  ^{3}+\tfrac{a_{4}}{4!}\left(
\bar{N}N\right)  ^{4},
\end{align}%
\begin{equation}
a_{0}=1,\qquad a_{1}=0,\qquad a_{2}=2c_{2},\qquad a_{3}=6c_{3},\qquad
a_{4}=12c_{2}^{2}+24c_{4}.
\end{equation}

We now to try find a real function $g(f)$ such that%
\begin{equation}
\int_{-\infty}^{\infty}df\,\exp\left[  f\bar{N}N+g(f)\right]  =a_{0}+a_{1}%
\bar{N}N+\tfrac{a_{2}}{2!}\left(  \bar{N}N\right)  ^{2}+\tfrac{a_{3}}%
{3!}\left(  \bar{N}N\right)  ^{3}+\tfrac{a_{4}}{4!}\left(  \bar{N}N\right)
^{4}.
\end{equation}
We observe that a Hubbard-Stratonovich transformation of this form maintains
the positive functional integral measure. \ Expanding the left-hand side, we
have%
\begin{equation}
a_{n}=\int_{-\infty}^{\infty}df\,f^{n}\exp\left[  g(f)\right]  ,\text{
\ }n=0,1,2,3,4.
\end{equation}
Finding sufficient and necessary conditions for the existence of $g(f)$ is
known in the mathematics literature as the truncated Hamburger moment problem.
\ This problem has been solved \cite{Adamyan:2003}\cite{Curto:1991}, and in
our case $g(f)$ exists if and only if the so-called block-Hankel matrix,%
\begin{equation}%
\begin{bmatrix}
a_{0} & a_{1} & a_{2}\\
a_{1} & a_{2} & a_{3}\\
a_{2} & a_{3} & a_{4}%
\end{bmatrix}
=%
\begin{bmatrix}
1 & 0 & 2c_{2}\\
0 & 2c_{2} & 6c_{3}\\
2c_{2} & 6c_{3} & 12c_{2}^{2}+24c_{4}%
\end{bmatrix}
,
\end{equation}
is positive semi-definite, with the added condition that if $c_{2}=0$ then
$c_{4}=0$. \ The determinant of this matrix is $16c_{2}^{3}-36c_{3}%
^{2}+48c_{2}c_{4}$. \ With an attractive two-nucleon force and small three and
four-nucleon forces, the conditions are clearly satisfied. \ Whether or not
these conditions are satisfied in the real world and at which lattice spacings
is beyond the scope of this letter. \ But hopefully this will be numerically
determined in the near future.

To summarize, the physics of low-energy symmetric nuclear matter is close to
the Wigner limit. \ We have proven that the energy of any three-nucleon state
is bounded below by the average energy of the lowest two-nucleon and
four-nucleon states. \ We have calculated the corrections due to the
lowest-order terms breaking Wigner symmetry and shown that the inequalities
are satisfied. \ We have proven general energy convexity results for $SU(N)$
and shown that all of these inequalities are satisfied for $N=4$. \ We have
also discussed the inclusion of Wigner-symmetric three and four-nucleon forces.

The authors thank John Thomas for discussions and Paulo Bedaque for his
hospitality at Lawrence Berkeley Laboratory where part of this research was
completed. \ This work was supported by the U.S. Department of Energy under
grants DE-FG-88ER40388 and DE-FG02-04ER41335 and the National Science Council
of Taiwan, ROC.

\bibliographystyle{h-physrev3}
\bibliography{NuclearMatter}

\end{document}